\documentclass[letterpaper, 10 pt, conference]{ieeeconf}  

\usepackage[T1]{fontenc}

\usepackage{amsmath}
\usepackage{algorithm}
\usepackage{algpseudocode}
\usepackage{bm}
\usepackage{cite}
\usepackage{amsmath,amssymb,amsfonts}
\usepackage{graphicx,color}
\usepackage{textcomp}
\def\BibTeX{{\rm B\kern-.05em{\sc i\kern-.025em b}\kern-.08em
    T\kern-.1667em\lower.7ex\hbox{E}\kern-.125emX}}
\AtBeginDocument{\definecolor{ojcolor}{cmyk}{0.93,0.59,0.15,0.02}}

\usepackage{graphicx}
\usepackage{soul,color}

\usepackage{colortbl} 

\definecolor{g}{rgb}{0.56, 0.93, 0.56}

\definecolor{bb}{rgb}{0.56, 0.86, 0.93}

\definecolor{rr}{rgb}{0.96, 0.06, 0.03}

\definecolor{vv}{rgb}{0.56, 0.06, 0.86}

\definecolor{pp}{rgb}{0.96, 0.76, 0.96}

\IEEEoverridecommandlockouts                              

\title{\LARGE \bf GPS Spoofing Attack Detection in Autonomous Vehicles Using Adaptive DBSCAN}

\author{Ahmad Mohammadi$^{1}$, Reza Ahmari$^{1}$, Vahid Hemmati$^{1}$, Frederick Owusu-Ambrose$^{1}$,\\Mahmoud Nabil Mahmoud$^{2}$, Parham Kebria$^{1}$, Abdollah Homaifar$^{1*}$, and Mehrdad Saif$^{3}$
\thanks{$^{1}$Authors are with the Department of Electrical and Computer Engineering at North Carolina A\&T State University, Greensboro, NC 27411, USA.}
\thanks{$^{2}$MN. Mahmoud is with the University of Alabama, Tuscaloosa, AL 35487, USA.}
\thanks{$^{3}$M. Saif is with the Department of Electrical and Computer Engineering, Windsor University, Windsor, ON N9B 3P4, Canada.}
\thanks{This research is supported by the University Transportation System (UTC), Department of Transportation, USA, through grant number 69A3552348327. Also, this work is partially supported by (the NASA-ULI under Cooperative Agreement Number 80NSSC20M0161.}
\thanks{$^{*}$Corresponding author: homaifar@ncat.edu}
}

\begin{document}

\maketitle
\thispagestyle{empty}
\pagestyle{empty}

\begin{abstract}

As autonomous vehicles become an essential component of modern transportation, they are increasingly vulnerable to threats such as GPS spoofing attacks. This study presents an adaptive detection approach utilizing a dynamically tuned Density Based Spatial Clustering of Applications with Noise (DBSCAN) algorithm, designed to adjust the detection threshold $(\varepsilon)$ in real-time. The threshold is updated based on the recursive mean and standard deviation of displacement errors between GPS and in-vehicle sensors data, but only at instances classified as non-anomalous. Furthermore, an initial threshold, determined from 120,000 clean data samples, ensures the capability to identify even subtle and gradual GPS spoofing attempts from the beginning. To assess the performance of the proposed method, five different subsets from the real-world Honda Research Institute Driving Dataset (HDD) are selected to simulate both  large and small magnitude GPS spoofing attacks. The modified algorithm effectively identifies turn-by-turn, stop, overshoot, and multiple small biased spoofing attacks, achieving detection accuracies of 98.62$\pm$1\%, 99.96$\pm$0.1\%, 99.88$\pm$0.1\%, and 98.38$\pm$0.1\%, respectively. This work provides a substantial advancement in enhancing the security and safety of AVs against GPS spoofing threats.

\end{abstract}

\section{INTRODUCTION}

\begin{figure}[htbp]
\centering
\includegraphics[width=0.9\columnwidth]{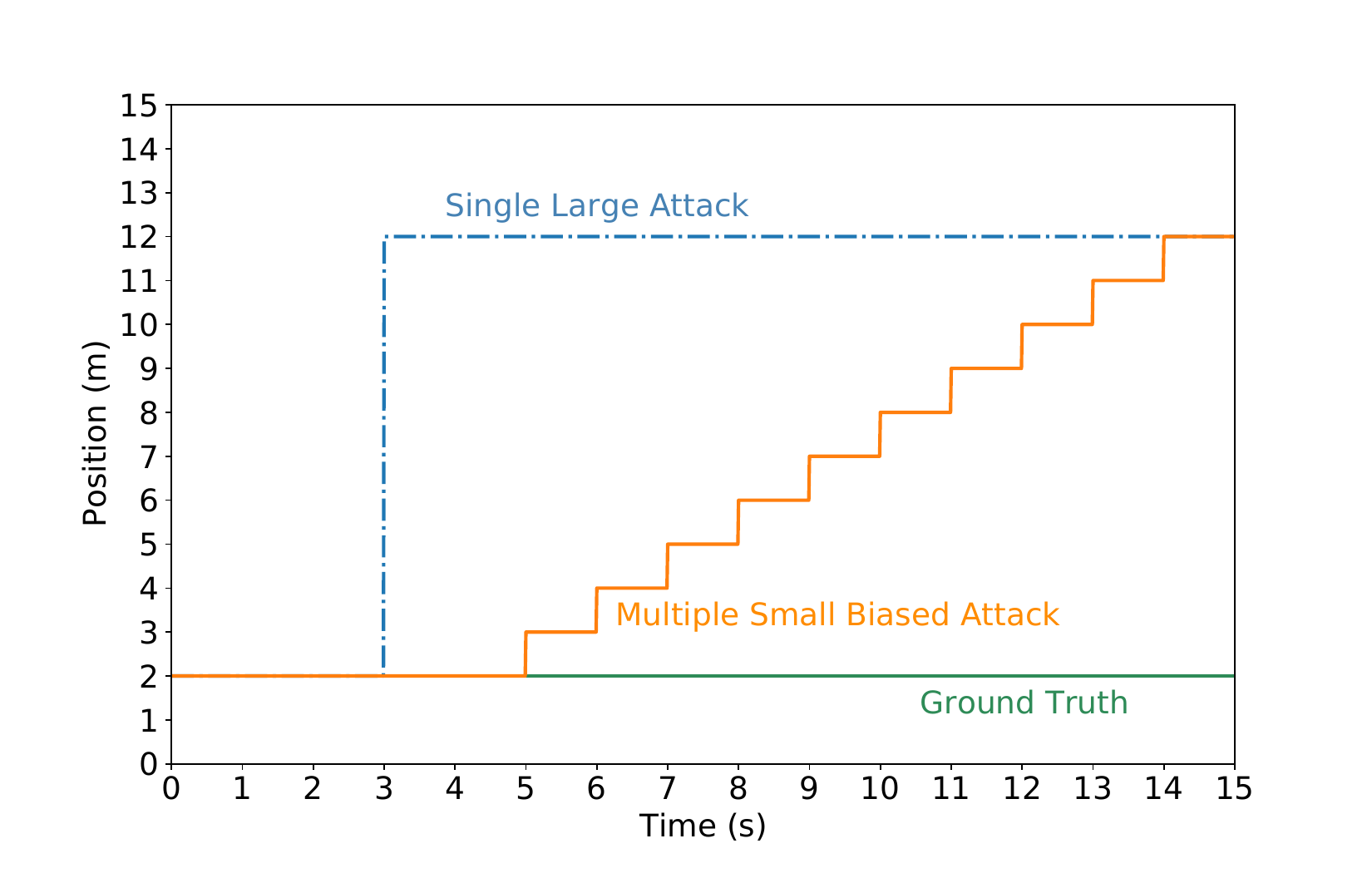}
\caption{Representation of a single attack with a magnitude of 10 m at t=3 second and a series of multiple small biased attacks applied in 10 steps (1 m in each step) to achieve the same attack magnitude.}
\label{fig:pic00}
\end{figure}

Autonomous vehicles (AVs) are rapidly transforming modern transportation by enhancing safety, efficiency, and convenience. However, this increasing dependence on automation introduces new security vulnerabilities, particularly adversarial threats targeting critical sensor systems. Since AVs rely on multiple sensors to interpret their surroundings and make navigation decisions \cite{deng2021deep, vahidmissionbased, 8755084}, they are susceptible to external manipulation. One of the most significant threats is Global Navigation Satellite System (GNSS) spoofing, where deceptive signals are transmitted to mislead an AV’s positioning system \cite{psiaki2016gnss, 7844895}. By distorting GNSS data, attackers can misrepresent an AV’s location, leading to incorrect navigation choices or hazardous driving maneuvers \cite{zeng2017practical, rezasmc, rezaifstrojan, 8755087, rezamynudd, rezaifsDD, frredifsa}. The consequences of such attacks range from minor route deviations to critical safety hazards, including collisions or vehicle takeover. GNSS spoofing attacks can occur in different forms. A turn-by-turn attack subtly alters the spoofed GNSS data over time, steering the AV off its intended path and potentially directing it toward hazardous areas or wrong destinations \cite{dasgupta2022sensor, chowdhury2024performance, van2018classification}. In a stop attack, a stationary vehicle is deceived into appearing as if it is moving, which may result in unsafe responses when it actually starts moving again \cite{dasgupta2022sensor, van2018classification}. Conversely, an overshoot attack makes a moving vehicle appear stationary, potentially triggering abrupt stops or unsafe maneuvers at intersections or roadway forks \cite{dasgupta2022sensor, van2018classification}. These sophisticated spoofing strategies can severely disrupt AV decision-making, endangering passengers, pedestrians, and other vehicles on the road \cite{dasgupta2022sensor}. Beyond the immediate safety risks, persistent GNSS spoofing can erode public trust in AV technology, hindering widespread adoption and further development. Despite notable progress in AV security, detecting advanced spoofing attacks remains a formidable challenge. Traditional anti-spoofing methods struggle with identifying attacks that involve gradual, stepwise location manipulations. As depicted in Figure \ref{fig:pic00}, a single large-scale spoofing event results in a sudden positional shift (blue dotted curve), which is easier to detect. In contrast, a more sophisticated attack involves multiple small biased steps that shift the vehicle’s perceived location incrementally in the same direction (red curve), making detection significantly more difficult. Conventional detection mechanisms often fail to identify these gradual distortions, allowing attackers to exploit AV vulnerabilities without immediate detection.

\begin{figure*}[htbp]
\centering
\includegraphics[width=0.7\textwidth]{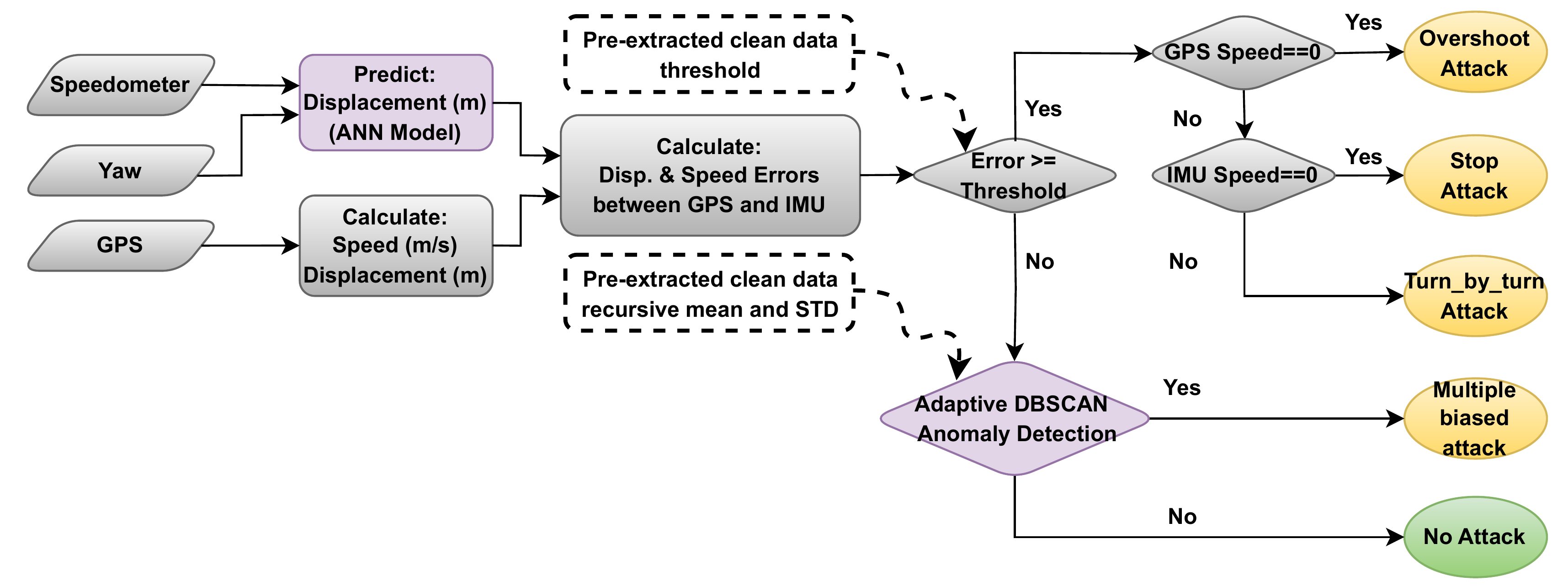}
\caption{The proposed framework for detecting GPS spoofing attacks: multiple small biased attacks are detected with adaptive DBSCAN by using recursive mean and STD calculation from the uncertainty in clean datasets.}
\label{fig:pic01}
\end{figure*}

This paper presents an innovative approach for detecting multiple small biased attacks using an adaptive DBSCAN algorithm. By real-time adaptation of clustering parameters, the proposed method successfully identifies both large-scale and small GPS spoofing attempts. The remainder of this paper is structured as follows: Section \ref{sec-related} reviews existing research in this domain and highlights key gaps in the literature. Section \ref{sec-methodology} elaborates on the proposed detection framework. Section \ref{sec-evaluation} presents the experimental results demonstrating the effectiveness of the spoofing attack detection method. Finally, Section \ref{sec-summary-future} summarizes the findings and discusses possible directions for future research.

\section{RELATED WORKS}\label{sec-related}
GNSS spoofing attack detection approaches fall into five categories: (i) encryption-based defenses, (ii) codeless cross-correlation techniques, (iii) signal feature analysis, (iv) antenna-based methods \cite{zidan2020gnss}, and (v) machine learning (ML) approaches.

Encryption-based strategies are used in military applications to safeguard GNSS signals. However, their dependence on cryptographic algorithms introduces substantial computational complexity, making them expensive for AVs implementations. Moreover, these methods often necessitate specialized hardware, which further escalates costs and limits their feasibility in commercial AV deployments.

Codeless cross-correlation techniques utilize the correlation of encrypted GPS L1 P(Y) signals among multiple receivers \cite{o2013real, o2010real}. Though effective in detecting spoofing, such techniques are hardware-intensive and computationally sophisticated and thus not readily implementable in large scale. With a rise in the number of receivers, there is a rise in computational complexity and cost in handling the signals.

Signal feature analysis identifies anomalies using observation of varied features of a signal, like received signal strength (RSS) \cite{yang2012detection}, spatial coherence \cite{daneshmand2012low}, pseudorange inconsistencies, time-of-arrival variations, and other features derived from a signal. It is still difficult to distinguish between normal variations in a signal and malicious spoofing, and in most instances, this results in false alarms or missed attacks. These approaches are also reliant on a pre-defined normal behavior baseline for a signal, which is difficult to define and maintain in dynamic environments for AV.

Antenna-based spoofing detection methods utilize multiple GNSS antennas and carrier-phase measurements to detect inconsistencies in signal propagation \cite{psiaki2014gnss}. While these techniques are effective, they demand substantial computational resources and involve intricate hardware setups, increasing both cost and system complexity \cite{zidan2020gnss, liu2021stars}. Another approach involves comparing acceleration data obtained from inertial measurement units (IMUs) with those estimated from GNSS signals \cite{neish2018uncoupled}. However, its effectiveness in AVs is limited due to their relatively restricted acceleration variations. Furthermore, certain detection strategies assess discrepancies between IMU-derived positioning and GNSS-reported locations to identify potential spoofing incidents \cite{tanil2018experimental, manickam2016using, dasgupta2022sensor}.

ML and deep learning (DL) methods have gained prominence in detecting GNSS spoofing attacks. Support Vector Machines (SVMs) have been applied to estimate the state of unmanned aerial vehicles (UAVs) for spoofing detection \cite{borhani2020deep}. Additionally, classifiers such as K-Nearest Neighbor (KNN) and Naïve Bayes have been used to analyze signal features, including early-late phase variations and power fluctuations, to identify spoofing events \cite{sun2017gps}. Deep learning models utilize in-vehicle sensor data to predict expected GNSS locations, enabling the detection of discrepancies that may indicate spoofing attempts \cite{dasgupta2022sensor}. Furthermore, ML techniques have been integrated into AVs for behavioral analysis, encompassing tasks such as misbehavior detection and predictive modeling \cite{AVcyberattacks, misbehaviordetection}.

Although significant progress has been made in anti-spoofing strategies \cite{kamal2021gps, dasgupta2022sensor, 10318172}, most existing research works have concentrated on detecting sudden spoofing attacks, often neglecting more gradual and sophisticated signal manipulations that happens in a step by step manner over time. A major challenge remains in identifying such spoofing attacks, where subtle and continuous shifts in positioning data make it difficult to distinguish between genuine and falsified signals. Addressing this gap is crucial for enhancing AV security against advanced GPS spoofing threats.

\section{PROPOSED METHODOLOGY}\label{sec-methodology}
This section outlines the detection methodology for both large-scale spoofing attacks—including turn-by-turn, stop, and overshoot attacks \cite{dasgupta2022sensor}— and multiple small biased attacks \cite{ICAIC2025}. The discussion is organized into the following two subsections:

\subsection{Detection of Single Large Magnitude Attacks}
Figure \ref{fig:pic01} illustrates the attack detection framework designed to identify both large-magnitude and small-magnitude spoofing attacks. A deep neural network (DNN) is employed to predict the vehicle's next-step displacement based on speedometer and yaw angle from the gyroscope \cite{ICAIC2025, ahmadifsa, ahmadvehicular}. Concurrently, the vehicle's displacement and speed are obtained from GPS data, and the difference between the predicted and measured values is computed. These errors are continuously monitored and compared against pre-extracted threshold values derived from clean data to detect anomalies in GPS signals. If the computed errors exceed the established thresholds, an anomaly is flagged, and the spoofing attack type is classified by analyzing speed readings from both data sources. Specifically, an overshoot attack is identified when the GPS speed is zero while the speedometer indicates motion. Conversely, a stop attack is detected when the speedometer reports that the vehicle is stationary while GPS speed is nonzero. Additionally, a turn-by-turn attack is recognized when both sources report nonzero speeds but discrepancies exist between the GPS-derived and in-vehicle sensor-based displacements. To develop the data-driven displacement prediction model, input features including speedometer readings, yaw angle $\psi(t)$, time intervals between samples, and GPS data are collected. GPS-based displacements, measured in meters, are computed every tenth data point as described in \eqref{eq:01} \cite{robusto1957cosine}.

{\small
\begin{equation}
    \begin{split}
        &disp_h=\\
        &2r\sin^{-1}\left(\sqrt{\sin^2\left(\frac{\Delta\phi}{2}\right)+\cos(\phi_1) \cos(\phi_2) \sin^2\left(\frac{\Delta\psi}{2}\right)}\right)
    \end{split}
    \label{eq:01}
\end{equation}
}
where:
{\small
\begin{align*}
    &disp_h \text{ is the distance between two points on the Earth's surface (m)} \\
    &r \text{ is the Earth's radius (6378000 m)} \\
    &\Delta\phi = \phi_2 - \phi_1 \text{ is the difference in latitudes (rad)} \\
    &\Delta\psi = \psi_2 - \psi_1 \text{ is the difference in longitudes (rad)}
\end{align*}
}

\begin{figure}[htbp]
\centering
\includegraphics[width=0.9\columnwidth]{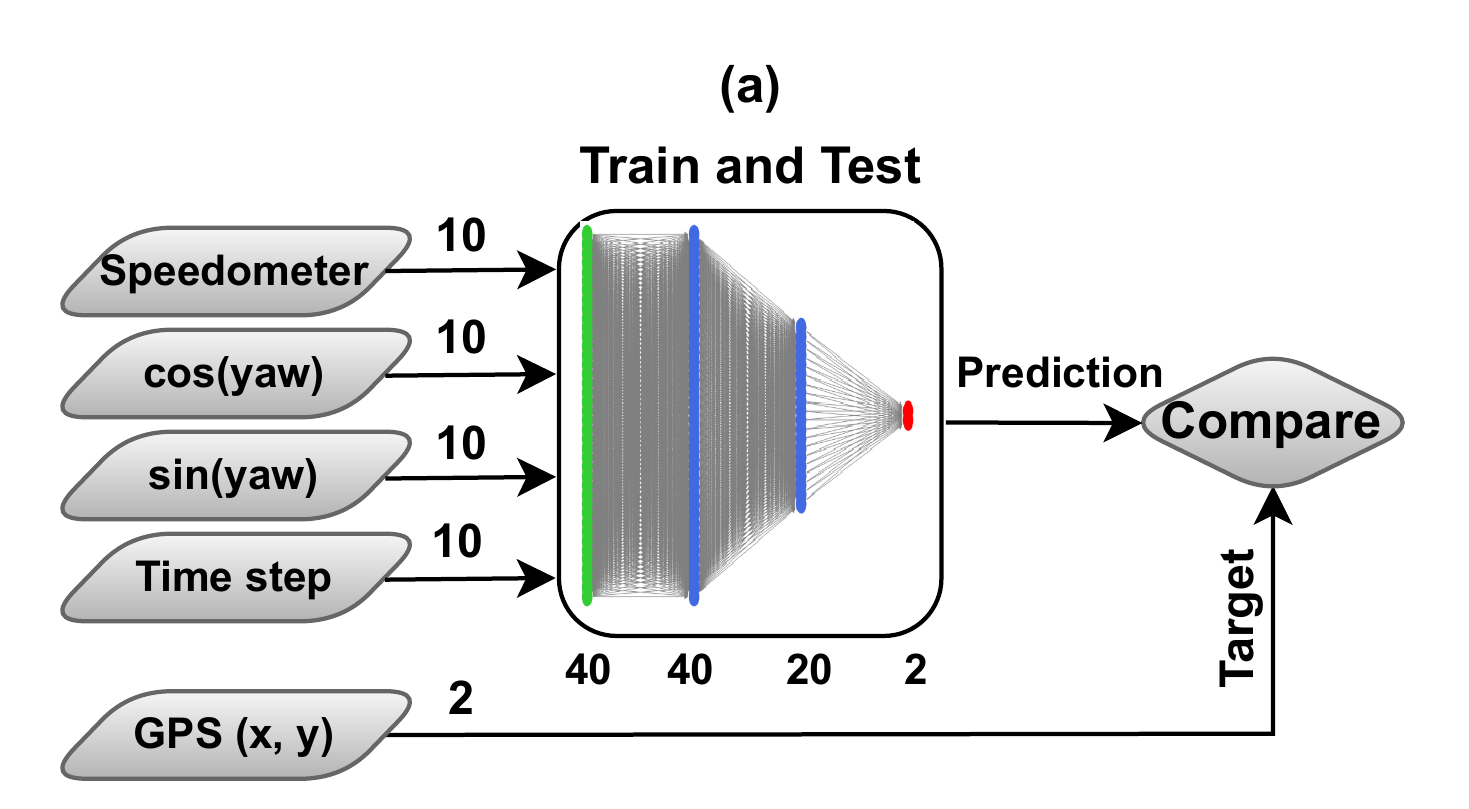}
\includegraphics[width=0.8\columnwidth]{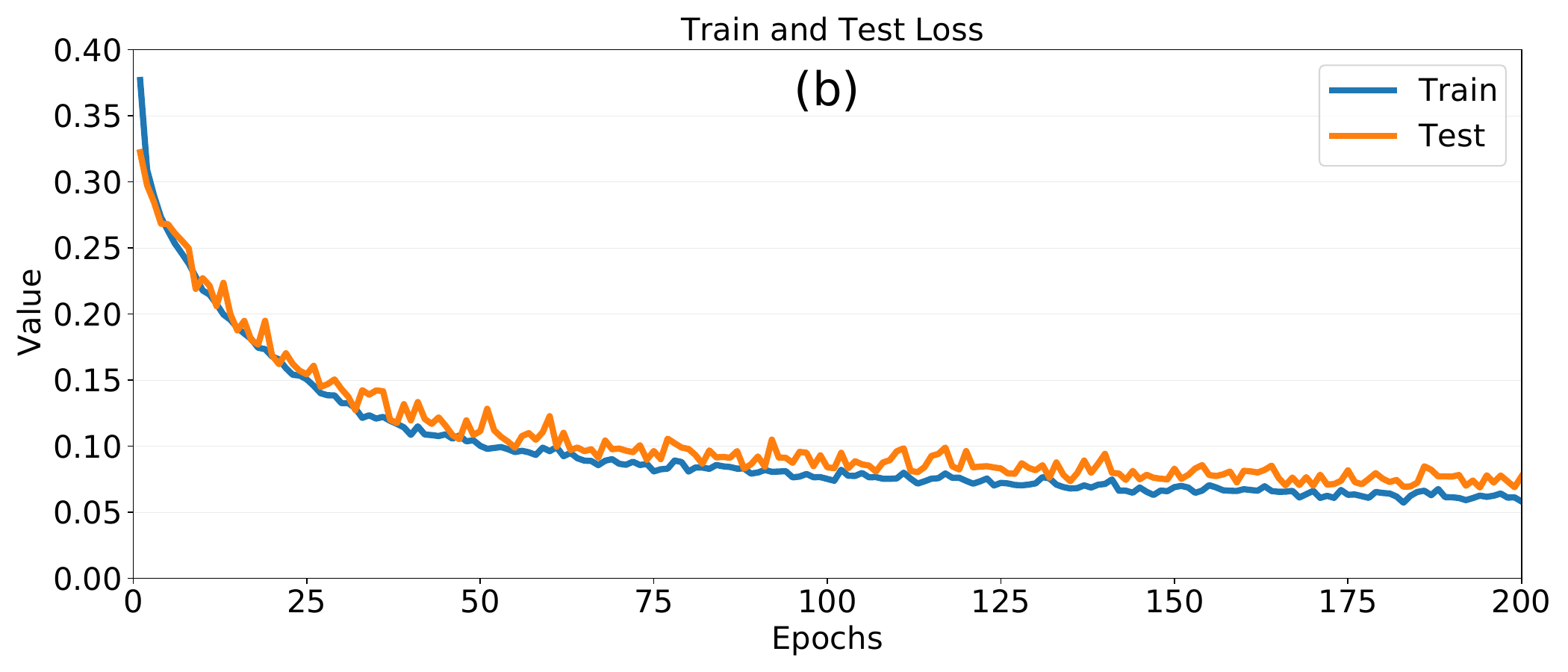}
\vspace{0.5em} 
\caption{(a) The proposed neural network receives ten consecutive samples of speed, \textbf{\(\cos(\psi)\)}, \textbf{\(\sin(\psi)\)}, and time steps as input, while the GPS-perceived displacement—sampled every tenth data point—serves as the target output. (b) The training and test loss curves, along with performance metrics, demonstrate a steady decline in error over 200 epochs, indicating a smooth learning process.}
\label{fig:pic02}
\end{figure}

 Since displacement does not have a direct linear relationship with $\psi(t)$, $\cos(\psi)$ and $\sin(\psi)$ are utilized as alternative inputs to the neural network to enhance the accuracy of displacement predictions along the x and y axes. Additionally, time intervals between consecutive data samples are computed and incorporated as an extra input feature. As depicted in Figure \ref{fig:pic02}(a), the neural network processes a total of 40 samples, 10 of each feature, including ten sequential samples of speed, $\cos(\psi)$, $\sin(\psi)$, and time intervals. GPS-derived displacements, extracted every ten samples, serve as target values for comparison against the neural network's predictions in both the x and y directions. Section \ref{sec-evaluation} delves into the development and fine-tuning of the ANN model. 

\subsection{Detection of Multiple Small Biased Attacks}
When a spoofing attack persists over multiple instances with magnitudes smaller than the system’s typical detection threshold, the risk becomes more severe \cite{ICAIC2025, ahmadifsa}. In such cases, although each individual attack remains minor and falls within acceptable error margins, their cumulative effect can substantially shift the vehicle's position without triggering an alert. This occurs because conventional detection mechanisms evaluate each data point independently against predefined thresholds. As a result, the vehicle may gradually drift off its intended trajectory without immediate detection. 
To address these subtle yet impactful attacks, we propose an enhanced algorithm that modifies the conventional DBSCAN \cite{DBSCAN}. DBSCAN is an unsupervised clustering algorithm \cite{elshamy2024fine, mahjourian2025sanitizing} that identifies dense regions of data points, distinguishing clusters from sparse areas that are considered noise. It utilizes two primary parameters: $\boldsymbol{\varepsilon}$, which specifies the neighborhood radius, and \textbf{min-Pts}, which determines the minimum number of points required to form a cluster. DBSCAN is particularly appropriate for identifying clusters of arbitrary shapes and handling noisy datasets. Our modified version of DBSCAN (see algorithm \ref{algorithm:1}) is able to analyze the displacement errors between the GPS and in-vehicle sensors in clean instances of data dynamically \cite{elahi2024matter} and in real-time and adjust its $\boldsymbol{\varepsilon}$ accordingly to have a dynamic threshold. The followings are the modifications done on the DBSCAN:
\begin{itemize}
    \item The adaptive DBSCAN detects a single primary cluster that represents the normal condition in the absence of an attack.
    \item Since there is no \textit{min-pts} parameter, any points located beyond $\varepsilon$ from the main cluster are classified as anomalies, regardless of their count.
    \item The $\varepsilon$ parameter is recursively updated with each sensor reading to efficiently track the error distribution and its recent trend to have a dynamic threshold for the DBSCAN.
    \item In order to develop an algorithm that detects anomalies in real-time, initial values of the mean and standard deviation (STD, $\sigma$) are required. Therefore, we ran the algorithm on a clean dataset with 120,000 rows, which is why the parameter $n$ is initially set to 120,000 for the recursive calculation of the mean and $\sigma$.

    \item According to the normal distribution, 99.73\% of points lie within $3\sigma$ around the mean value. We have chosen $5\sigma$ as the dynamic threshold $\varepsilon$ to account for previous attacks or anomalies that were possibly missed.
\end{itemize}




    

\begin{algorithm}
\caption{Adaptive DBSCAN with dynamic threshold ($\varepsilon$)}
\begin{algorithmic}[1]
\Require initial mean ($\mu_0$) and standard deviation ($\sigma_0$)
\Ensure anomaly flag, updated recursive mean ($\mu$) and standard deviation ($\sigma$)

\State initialize $n, \mu, \sigma$

\While{data is coming}
    \State $anomaly \gets False$
    \State $deviation \gets |x - \mu|$ \Comment{Deviation from the recursive mean value}
    \State $\varepsilon \gets 5\sigma$

    \If{$-\varepsilon \leq deviation \leq \varepsilon$}
        \State $n \gets n + 1$
        \State $\mu \gets \mu + \frac{(x - \mu)}{n}$
        \State $\sigma \gets \sqrt{\frac{(n - 1) \cdot \sigma^2 + (x - \mu) \cdot (x - \mu)}{n}}$
    \Else
        \State $anomaly \gets True$
        
    \EndIf
    \State \Return anomaly, mean$(\mu)$, STD$(\sigma)$

\EndWhile

\end{algorithmic}
\label{algorithm:1}
\end{algorithm}

 This approach is represented as adaptive DBSCAN in figure \ref{fig:pic01}. If the error between GPS and in-vehicle sensors is less than the static thresholds, the adaptive DBSCAN detects anomalies caused by multiple small biased attacks using its dynamic $\varepsilon$.
 The results of evaluating this methodology, including accuracy, Specificity, sensitivity and area under curve (AUC) are discussed in section \ref{sec-evaluation}.

\section{EVALUATION AND RESULTS}\label{sec-evaluation}
This research leverages the Honda Research Institute Driving Dataset (HDD) \cite{ramanishka2018CVPR} to develop and assess a resilient framework for detecting GNSS spoofing attacks. The dataset encompasses different sensor data collected from a commercially available vehicle, integrating information from cameras, LiDAR, GNSS, an inertial measurement unit (IMU), and the controller area network (CAN) bus. 

Data acquisition took place in diverse driving conditions, including suburban neighborhoods, high-density urban roads, and highway routes within the San Francisco Bay Area. Given that AVs are equipped with a comparable array of sensors, the HDD serves as an ideal benchmark for generating and evaluating GNSS spoofing attack scenarios specifically tailored to AV applications. 

The dataset provides detailed sensor measurements from the vehicle's CAN bus, recorded at a sampling rate of 100 Hz. These include parameters such as relative accelerator pedal position (expressed as a percentage), steering wheel angle (measured in degrees), steering wheel rotational speed (degrees per second), vehicle velocity (kilometers per second), brake pressure (kilopascals), and yaw rate (degrees per second). Additionally, GNSS data is captured at the same frequency using an Automotive Dynamic Motion Analyzer (ADMA) manufactured by Genesys Electronic GmbH. The ADMA is integrated with a differential global positioning system (DGPS) to enhance positioning accuracy, ensuring precise localization data for further analysis.

\subsection{ANN Model Development and the Detection of Single Large Magnitude Attacks}
\subsubsection{Model Development}
As outlined in Section \ref{sec-methodology} (see Figure \ref{fig:pic02} (a)), the artificial neural network (ANN) architecture consists of an input layer, two hidden layers, and an output layer, with 40, 40, 20, and 2 neurons, respectively. The input and hidden layers utilize the \textit{tanh} activation function, while the output layer employs a linear activation function. Table \ref{table:1} summarizes the model's hyperparameters.

The model is trained and evaluated using a dataset containing 120,000 samples, which were divided into training and testing sets using a 70:30 split ratio. This split balances model training and evaluation, helping prevent overfitting. Training is done for 200 epochs using the Adam optimizer with a batch size of 32. Throughout training, the loss function—Mean Absolute Error (MAE)—consistently decreased and the close alignment of training and testing loss curves indicates effective learning, as illustrated in Figure \ref{fig:pic02}(b). Training was concluded at 200 epochs since no significant reduction in the loss function was observed beyond this point, except for minor fluctuations.

\begin{table}[h!]
\caption{Hyperparameters of the developed displacement predictor model.}
\centering
\begin{tabular}{|l|c|}
\hline
\rowcolor[gray]{0.8} 
\textbf{Hyperparameters}       & \textbf{Value} \\
\hline
Number of neurons (1\textsuperscript{st} and 2\textsuperscript{nd} layers) & 40 \\
Number of neurons (3\textsuperscript{rd} layer) & 20 \\
Number of neurons (4\textsuperscript{th} layer) & 2 \\
Number of epochs                      & 200 \\
Batch size                            & 32 \\
Activation function (layers 1-3)      & tanh \\
Activation function (output layer)    & linear \\
Optimizer                             & Adam \\
\hline
\end{tabular}
\label{table:1}
\end{table}

\subsubsection{Attack Detection}
To assess the effectiveness of the proposed attack detection approach presented in Figure \ref{fig:pic01}, the framework is evaluated using five distinct scenarios from HDD. These scenarios are carefully selected to encompass a range of driving conditions, including both urban and suburban environments with varying speed profiles.

The simulated attack scenarios include turn-by-turn, overshoot, and stop attacks, ensuring realistic representations in terms of attack magnitude, duration, and vehicle speed. For instance, during a turn-by-turn or overshoot attack, the vehicle must be in motion, whereas in a stop attack, the vehicle remains stationary. Additionally, for overshoot and stop attacks, GPS data is manipulated in a manner that prevents abrupt changes in GPS-reported speed, making the attack more covert and difficult to detect.

The algorithm's performance is then evaluated based on these simulated attack scenarios. Furthermore, each type of attack is executed five times per dataset with randomized attack magnitudes, durations, and start/end times. This ensures that the proposed framework is rigorously tested for robustness and effectiveness across various attack conditions.

\textbf{First}, clean data subsets are utilized as inputs to the algorithm to establish displacement and speed thresholds. The algorithm extracts the maximum observed error between GPS data and in-vehicle sensors, selecting these values as thresholds to account for sensor uncertainty. This approach ensures that anomalies remain detectable despite inherent sensor inaccuracies, thereby reducing false negatives—cases where an attack is present but goes undetected. Under clean conditions, the displacement and speed thresholds are determined to be 1.79 meters and 2.91 meters per second.

\textbf{Second}, for each category of single large-magnitude attacks—turn-by-turn, stop, and overshoot—a total of twenty-five attack instances are generated. This is achieved by introducing five attack scenarios per clean data subset, each with randomly assigned magnitudes and durations. As a result, the data comprises a combination of compromised and uncompromised data. The mean and STD of accuracy, sensitivity, specificity, and AUC for detecting turn-by-turn, stop, and overshoot attacks are summarized in Table \ref{table:2}.

\subsection{Detection of Multiple Small Biased Attacks with Magnitudes Smaller than System Threshold}

\textbf{First}, to facilitate real-time detection of multiple small biased attacks, the initial mean and standard deviation (STD) are recursively computed from a clean dataset containing 120k entries. These statistical parameters are then utilized by the adaptive DBSCAN algorithm for dynamic thresholding.

\textbf{Second}, to evaluate the effectiveness of the proposed adaptive DBSCAN algorithm in detecting multiple small biased attacks, the GPS data is systematically manipulated by incrementally shifting the x and y coordinates by one meter over 101 consecutive steps. This results in a cumulative displacement of approximately 141 meters, with each step introducing a shift of 1.41 meters—remaining within the system’s detection threshold of 1.79 meters. The experiment is repeated five times on each clean dataset, leading to a total of 25 distinct attack scenarios. The mean and standard deviation (STD) of accuracy, sensitivity, specificity, and AUC for detecting these attacks are summarized in Table \ref{table:2}.

\begin{table}[h!]
\caption{Summary of results for all four types of spoofing attack detection. All the numbers are reported in percent.}
\centering
\begin{tabular}{|l|c|c|c|c|}
\hline
\rowcolor[gray]{0.8} 
\textbf{Attack Type} & \textbf{Accuracy}           & \textbf{Sensitivity} & \textbf{Specificity} & \textbf{AUC} \\
\hline
Turn-by-turn         & 98.62$\pm$1         & 100       & 98.62$\pm$1.1       & 99.3$\pm$0.6 \\
Stop                 & 99.96$\pm$0.1     & 98.9$\pm$1.8       & 100       & 99.44$\pm$0.9 \\
Overshoot            & 99.88$\pm$0.1       & 99.16$\pm$0.2       & 100       & 99.63$\pm$0.5 \\
Mul. Biased & 98.38$\pm$0.1        & 100 & 99.18$\pm$1.1 & 99.3$\pm$0.6 \\
\hline
\end{tabular}
\label{table:2}
\end{table}

\section{CONCLUSIONS AND FUTURE WORK}\label{sec-summary-future}
This paper presents a Data-Driven framework combined with an adaptive DBSCAN algorithm for detecting GPS spoofing attacks by leveraging GPS and in-vehicle sensor data. The Data-Driven model predicts the vehicle's displacement in subsequent time steps. These predicted displacements are then compared with those obtained from GPS data, and deviations are assessed against predefined thresholds derived from clean data to detect anomalies in real-time. Additionally, different types of attacks are classified by continuously monitoring speed measurements from both GPS and the IMU. Beyond traditional anomaly detection, the modified adaptive DBSCAN algorithm enhances the framework's capability by identifying multiple small biased attacks, even when their magnitude remains below the system's detection threshold. This is achieved by dynamically adjusting the threshold parameter ($\varepsilon$) based on real-time displacement errors between GPS and in-vehicle sensors in clean data instances. In summary, the integration of the Data-Driven model and the adaptive DBSCAN algorithm significantly improves accuracy and efficiency in detecting various types of GPS spoofing attacks, including turn-by-turn, stop, overshoot, and multiple small biased attacks. The proposed approach achieves detection accuracies of 98.62$\pm$1\%, 99.96$\pm$0.1\%, 99.88$\pm$0.1\%, and 98.38$\pm$0.1\%, respectively, demonstrating its reliability and effectiveness in real-time applications.

Future research can further enhance GPS spoofing detection by integrating reinforcement learning (RL) techniques to dynamically adapt detection thresholds. Unlike traditional static thresholds, RL-driven approaches can continuously learn and optimize the thresholds based on evolving attack patterns and environmental conditions. By leveraging deep reinforcement learning (DRL) frameworks, the system can autonomously adjust its sensitivity to spoofing attacks, reducing false positives while maintaining high detection accuracy. Incorporating RL for adaptive thresholding would enable a more resilient and intelligent detection framework, further strengthening AV security against GPS spoofing threats.



\bibliographystyle{IEEEtran}
\bibliography{ref}

\begin{thebibliography}{10}
\providecommand{\url}[1]{#1}
\csname url@samestyle\endcsname
\providecommand{\newblock}{\relax}
\providecommand{\bibinfo}[2]{#2}
\providecommand{\BIBentrySTDinterwordspacing}{\spaceskip=0pt\relax}
\providecommand{\BIBentryALTinterwordstretchfactor}{4}
\providecommand{\BIBentryALTinterwordspacing}{\spaceskip=\fontdimen2\font plus
\BIBentryALTinterwordstretchfactor\fontdimen3\font minus \fontdimen4\font\relax}
\providecommand{\BIBforeignlanguage}[2]{{%
\expandafter\ifx\csname l@#1\endcsname\relax
\typeout{** WARNING: IEEEtran.bst: No hyphenation pattern has been}%
\typeout{** loaded for the language `#1'. Using the pattern for}%
\typeout{** the default language instead.}%
\else
\language=\csname l@#1\endcsname
\fi
#2}}
\providecommand{\BIBdecl}{\relax}
\BIBdecl

\bibitem{deng2021deep}
Y.~Deng, T.~Zhang, G.~Lou, X.~Zheng, J.~Jin, and Q.-L. Han, ``Deep learning-based autonomous driving systems: A survey of attacks and defenses,'' \emph{IEEE Transactions on Industrial Informatics}, vol.~17, no.~12, pp. 7897--7912, 2021.

\bibitem{vahidmissionbased}
V.~Hemmati, M.~Behnia, A.~Mohammadi, A.-R. Nuhu, and A.~Homaifar, ``Mission-based quadcopter flight simulation,'' in \emph{2024 AIAA DATC/IEEE 43rd Digital Avionics Systems Conference (DASC)}, 2024, pp. 1--7.

\bibitem{8755084}
P.~M. Kebria, R.~Alizadehsani, S.~M. Salaken, I.~Hossain, A.~Khosravi, D.~Kabir, A.~Koohestani, H.~Asadi, S.~Nahavandi, E.~Tunsel, and M.~Saif, ``Evaluating architecture impacts on deep imitation learning performance for autonomous driving,'' in \emph{2019 IEEE International Conference on Industrial Technology (ICIT)}, 2019, pp. 865--870.

\bibitem{psiaki2016gnss}
M.~L. Psiaki and T.~E. Humphreys, ``Gnss spoofing and detection,'' \emph{Proceedings of the IEEE}, vol. 104, no.~6, pp. 1258--1270, 2016.

\bibitem{7844895}
P.~M. Kebria, H.~Abdi, and S.~Nahavandi, ``Development and evaluation of a symbolic modelling tool for serial manipulators with any number of degrees of freedom,'' in \emph{2016 IEEE International Conference on Systems, Man, and Cybernetics (SMC)}, 2016, pp. 004\,223--004\,228.

\bibitem{zeng2017practical}
K.~C. Zeng, Y.~Shu, S.~Liu, Y.~Dou, and Y.~Yang, ``A practical gps location spoofing attack in road navigation scenario,'' in \emph{Proceedings of the 18th international workshop on mobile computing systems and applications}, 2017, pp. 85--90.

\bibitem{rezasmc}
R.~Ahmari, A.~Mohammadi, V.~Hemmati, M.~Mohammed, M.~N. Mahmoud, P.~Kebria, and A.~Homaifar, ``An experimental study of trojan vulnerabilities in uav autonomous landing,'' in \emph{2025 IEEE International Conference on Systems, Man, and Cybernetics (SMC) (accepted)}.\hskip 1em plus 0.5em minus 0.4em\relax IEEE, 2025.

\bibitem{rezaifstrojan}
R.~Ahmari, V.~Hemmati, A.~Mohammadi, M.~Mynuddin, P.~Kebria, M.~Mahmoud, and A.~Homaifar, ``Evaluating trojan attack vulnerabilities in autonomous landing systems for urban air mobility,'' \emph{Automation, Robotics \& Communications for Industry 4.0/5.0}, p.~80, 2025.

\bibitem{8755087}
P.~M. Kebria, A.~Khosravi, S.~Nahavandi, A.~Homaifar, and M.~Saif, ``Experimental comparison study on joint and cartesian space control schemes for a teleoperation system under time-varying delay,'' in \emph{2019 IEEE International Conference on Industrial Technology (ICIT)}, 2019, pp. 108--113.

\bibitem{rezamynudd}
M.~Mynuddin, S.~U. Khan, R.~Ahmari, L.~Landivar, M.~N. Mahmoud, and A.~Homaifar, ``Trojan attack and defense for deep learning-based navigation systems of unmanned aerial vehicles,'' \emph{IEEE Access}, vol.~12, pp. 89\,887--89\,907, 2024.

\bibitem{rezaifsDD}
R.~Ahmari, V.~Hemmati, A.~Mohammadi, P.~Kebria, M.~Mahmoud, and A.~Homaifar, ``A data-driven approach for uav-ugv integration,'' \emph{Automation, Robotics \& Communications for Industry 4.0/5.0}, p.~77, 2025.

\bibitem{frredifsa}
F.~Owusu-Ambrose, A.-r. Nuhu, B.~Lartey, A.~Mohammadi, and A.~Homaifar, ``Fuzzy equivalence relation clustering framework for distributed denial of service detection accuracy estimation,'' in \emph{IFSA Winter Conference on Automation, Robotics \& Communications for Industry 4.0/5.0 (ARCI)}, 2025.

\bibitem{dasgupta2022sensor}
S.~Dasgupta, M.~Rahman, M.~Islam, and M.~Chowdhury, ``A sensor fusion-based gnss spoofing attack detection framework for autonomous vehicles,'' \emph{IEEE Transactions on Intelligent Transportation Systems}, vol.~23, no.~12, pp. 23\,559--23\,572, 2022.

\bibitem{chowdhury2024performance}
Z.~U. Chowdhury, A.~R. Chowdhury, A.~Al~Jawad, R.~Murshed, A.~Rashid, M.~Mynuddin, R.~Ahmari, and A.~Mohammadi, ``Performance comparison of yolo models for safety helmet detection: Insights from yolov5 to yolov10 with transfer learning,'' \emph{Authorea Preprints}, 2024.

\bibitem{van2018classification}
J.~R. Van Der~Merwe, X.~Zubizarreta, I.~Luk{\v{c}}in, A.~R{\"u}gamer, and W.~Felber, ``Classification of spoofing attack types,'' in \emph{2018 European Navigation Conference (ENC)}.\hskip 1em plus 0.5em minus 0.4em\relax IEEE, 2018, pp. 91--99.

\bibitem{zidan2020gnss}
J.~Zidan, E.~I. Adegoke, E.~Kampert, S.~A. Birrell, C.~R. Ford, and M.~D. Higgins, ``Gnss vulnerabilities and existing solutions: A review of the literature,'' \emph{IEEE Access}, vol.~9, pp. 153\,960--153\,976, 2020.

\bibitem{o2013real}
B.~W. O'Hanlon, M.~L. Psiaki, J.~A. Bhatti, D.~P. Shepard, and T.~E. Humphreys, ``Real-time gps spoofing detection via correlation of encrypted signals,'' \emph{Navigation}, vol.~60, no.~4, pp. 267--278, 2013.

\bibitem{o2010real}
B.~W. O'Hanlon, M.~L. Psiaki, T.~E. Humphreys, and J.~A. Bhatti, ``Real-time spoofing detection in a narrow-band civil gps receiver,'' in \emph{Proceedings of the 23rd International Technical Meeting of the Satellite Division of The Institute of Navigation (ION GNSS 2010)}, 2010, pp. 2211--2220.

\bibitem{yang2012detection}
J.~Yang, Y.~Chen, W.~Trappe, and J.~Cheng, ``Detection and localization of multiple spoofing attackers in wireless networks,'' \emph{IEEE Transactions on Parallel and Distributed systems}, vol.~24, no.~1, pp. 44--58, 2012.

\bibitem{daneshmand2012low}
S.~Daneshmand, A.~Jafarnia-Jahromi, A.~Broumandon, and G.~Lachapelle, ``A low-complexity gps anti-spoofing method using a multi-antenna array,'' in \emph{Proceedings of the 25th international technical meeting of the satellite division of the institute of navigation (ION GNSS 2012)}, 2012, pp. 1233--1243.

\bibitem{psiaki2014gnss}
M.~L. Psiaki, B.~W. OHanlon, S.~P. Powell, J.~A. Bhatti, K.~D. Wesson, and T.~E. Schofield, ``Gnss spoofing detection using two-antenna differential carrier phase,'' in \emph{Proceedings of the 27th international technical meeting of the satellite division of the Institute of Navigation (ION GNSS+ 2014)}, 2014, pp. 2776--2800.

\bibitem{liu2021stars}
S.~Liu, X.~Cheng, H.~Yang, Y.~Shu, X.~Weng, P.~Guo, K.~C. Zeng, G.~Wang, and Y.~Yang, ``Stars can tell: a robust method to defend against $\{$GPS$\}$ spoofing attacks using off-the-shelf chipset,'' in \emph{30th USENIX Security Symposium (USENIX Security 21)}, 2021, pp. 3935--3952.

\bibitem{neish2018uncoupled}
A.~Neish, S.~Lo, Y.-H. Chen, and P.~Enge, ``Uncoupled accelerometer based gnss spoof detection for automobiles using statistic and wavelet based tests,'' in \emph{Proceedings of the 31st International Technical Meeting of the Satellite Division of The Institute of Navigation (ION GNSS+ 2018)}, 2018, pp. 2938--2962.

\bibitem{tanil2018experimental}
C.~Tanil, P.~M. Jimenez, M.~Raveloharison, B.~Kujur, S.~Khanafseh, and B.~Pervan, ``Experimental validation of ins monitor against gnss spoofing,'' in \emph{Proceedings of the 31st International Technical Meeting of the Satellite Division of The Institute of Navigation (ION GNSS+ 2018)}, 2018, pp. 2923--2937.

\bibitem{manickam2016using}
S.~Manickam and K.~O'Keefe, ``Using tactical and mems grade ins to protect against gnss spoofing in automotive applications,'' in \emph{Proceedings of the 29th International Technical Meeting of the Satellite Division of the Institute of Navigation (ION GNSS+ 2016)}, 2016, pp. 2991--3001.

\bibitem{borhani2020deep}
P.~Borhani-Darian, H.~Li, P.~Wu, and P.~Closas, ``Deep neural network approach to detect gnss spoofing attacks,'' in \emph{Proceedings of the 33rd International Technical Meeting of the Satellite Division of The Institute of Navigation (ION GNSS+ 2020)}, 2020, pp. 3241--3252.

\bibitem{sun2017gps}
M.~Sun, Y.~Qin, J.~Bao, and X.~Yu, ``Gps spoofing detection based on decision fusion with a k-out-of-n rule.'' \emph{Int. J. Netw. Secur.}, vol.~19, no.~5, pp. 670--674, 2017.

\bibitem{AVcyberattacks}
M.~Girdhar, J.~Hong, and J.~Moore, ``Cybersecurity of autonomous vehicles: A systematic literature review of adversarial attacks and defense models,'' \emph{IEEE Open Journal of Vehicular Technology}, vol.~4, pp. 417--437, 2023.

\bibitem{misbehaviordetection}
A.~Sharma and A.~Jaekel, ``Machine learning based misbehaviour detection in vanet using consecutive bsm approach,'' \emph{IEEE Open Journal of Vehicular Technology}, vol.~3, pp. 1--14, 2022.

\bibitem{kamal2021gps}
M.~Kamal, A.~Barua, C.~Vitale, C.~Laoudias, and G.~Ellinas, ``Gps location spoofing attack detection for enhancing the security of autonomous vehicles,'' in \emph{2021 IEEE 94th Vehicular Technology Conference (VTC2021-Fall)}.\hskip 1em plus 0.5em minus 0.4em\relax IEEE, 2021, pp. 1--7.

\bibitem{10318172}
N.~Niknejad and H.~Modares, ``Physics-informed data-driven safe and optimal control design,'' \emph{IEEE Control Systems Letters}, vol.~8, pp. 285--290, 2024.

\bibitem{ICAIC2025}
A.~Mohammadi, V.~Hemmati, R.~Ahmari, F.~Owusu-Ambrose, M.~N. Mahmoud, and A.~Homaifar, ``Detection of multiple small biased gps spoofing attacks on autonomous vehicles,'' in \emph{2025 IEEE 4th International Conference on AI in Cybersecurity (ICAIC)}, 2025, pp. 1--9.

\bibitem{ahmadifsa}
A.~Mohammadi, V.~Hemmati, R.~Ahmari, F.~Owusu-Ambrose, M.~Mahmoud, and A.~Homaifar, ``Gps spoofing attack detection on autonomous vehicles using modified dbscan with dynamic threshold,'' in \emph{Proceedings of the 5th IFSA Winter Conference on Automation, Robotics \& Communications for Industry 4.0/5.0 (ARCI’2025)}, 2025, pp. 74--76.

\bibitem{ahmadvehicular}
A.~Mohammadi, R.~Ahmari, V.~Hemmati, F.~Owusu-Ambrose, M.~N. Mahmoud, P.~Kebria, and A.~Homaifar, ``Detection of multiple small biased gps spoofing attacks on autonomous vehicles using time series analysis,'' \emph{IEEE Open Journal of Vehicular Technology}, pp. 1--13, 2025.

\bibitem{robusto1957cosine}
C.~C. Robusto, ``The cosine-haversine formula,'' \emph{The American Mathematical Monthly}, vol.~64, no.~1, pp. 38--40, 1957.

\bibitem{DBSCAN}
X.~Xu, M.~Ester, H.-P. Kriegel, and J.~Sander, ``A distribution-based clustering algorithm for mining in large spatial databases,'' in \emph{Proceedings 14th International Conference on Data Engineering}, 1998, pp. 324--331.

\bibitem{elshamy2024fine}
M.~R. Elshamy, M.~Elahi, A.~Patooghy, and A.-H.~A. Badawy, ``Fine-grained clustering-based power identification for multicores,'' in \emph{2024 IEEE 15th International Green and Sustainable Computing Conference (IGSC)}.\hskip 1em plus 0.5em minus 0.4em\relax IEEE, 2024, pp. 165--170.

\bibitem{mahjourian2025sanitizing}
N.~Mahjourian and V.~Nguyen, ``Sanitizing manufacturing dataset labels using vision-language models,'' \emph{arXiv preprint arXiv:2506.23465}, 2025.

\bibitem{elahi2024matter}
M.~Elahi, M.~R. Elshamy, A.-H. Badawy, M.~Fazeli, and A.~Patooghy, ``Matter: Multi-stage adaptive thermal trojan for efficiency \& resilience degradation,'' \emph{arXiv preprint arXiv:2412.00226}, 2024.

\bibitem{ramanishka2018CVPR}
V.~Ramanishka, Y.-T. Chen, T.~Misu, and K.~Saenko, ``Toward driving scene understanding: A dataset for learning driver behavior and causal reasoning,'' in \emph{Conference on Computer Vision and Pattern Recognition (CVPR)}, 2018.

\end{thebibliography}

\end{document}